\documentclass[12pt]{article}

\def\beq{\begin{eqnarray}}
\def\eeq{\end{eqnarray}}

\def\L*{{\cal L}_*}
\def\lsim{\mathrel{\rlap{\lower3pt\hbox{\hskip0pt$\sim$}}
     \raise1pt\hbox{$<$}}}         
\def\gsim{\mathrel{\rlap{\lower4pt\hbox{\hskip1pt$\sim$}}
     \raise1pt\hbox{$>$}}}         

\usepackage{amsmath}
\usepackage{amsfonts}

\begin{document}
\begin{titlepage}

\centerline{\Large \bf Tachyon-Free Non-Supersymmetric Strings on Orbifolds}
\medskip

\centerline{\large Zurab Kakushadze}

\bigskip

\centerline{\em 200 Rector Place, Apt 41F, New York, NY 10280}
\centerline{\tt zura@kakushadze.com}
\centerline{(November 26, 2007)}

\bigskip
\medskip

\begin{abstract}
{}We discuss tachyon-free examples of (Type IIB on) non-compact non-supersymmetric orbifolds. Tachyons are projected
out by discrete torsion between orbifold twists, while supersymmetry is broken by a Scherk-Schwarz phase ($+1$/$-1$ when acting on space-time bosons/fermions) accompanying some even order twists. The absence of tachyons is encouraging for constructing non-supersymmetric D3-brane gauge theories with stable infrared fixed points. The D3-brane gauge theories in our orbifold backgrounds have chiral ${\cal N} = 1$ supersymmetric spectra, but non-supersymmetric interactions.
\end{abstract}
\end{titlepage}

\newpage

\section{Introduction and Summary}\label{Intro}

{}In string theory supersymmetry breaking is a key ingredient for making a connection with nature. This applies to
string theory as a fundamental unified theory, as well as a model for describing strong interactions. One possibility is to have
dynamical supersymmetry breaking in a background with tree-level supersymmetry. Another is to construct backgrounds without
space-time supersymmetry already at tree-level.

{}Construction of tachyon-free non-supersymmetric backgrounds has a long history starting with a heterotic model in 10 dimensions
\cite{DH}, which is an example of an orbifold construction. The orbifold action is such that it projects out all
gravitinos in the untwisted sectors, while not reintroducing them in the twisted sectors. In general, the latter may contain tachyons. Here an important
distinction arises.

{}If some spatial directions are compactified, tachyons in the twisted sectors can be avoided by including shifts in the compact
directions. An example is the Scherk-Schwarz breaking \cite{SS}, where a half-shift around a circle
is accompanied by a supersymmetry breaking phase. The twisted ground state energy depends on
the radius of the circle, and for small radius becomes negative, so a tachyon appears. At large radius supersymmetry is restored.

{}It is more intricate to construct {\em non-compact} tachyon-free non-supersymmetric models. In the heterotic case
one can accompany a supersymmetry breaking twist by a twist or a shift in the internal (non-geometric) directions on
the bosonic side, thereby avoiding tachyons. In the case of Type II there are no internal directions, and in general
non-supersymmetric twisted sectors give rise to tachyons. If fact, there even appears to be folklore that
non-compact tachyon-free non-supersymmetric Type II orbifolds cannot be constructed.

{}The goal of this paper is to present a construction of non-compact tachyon-free non-supersymmetric Type II orbifolds. To the best of our
knowledge, the examples herein are first of the kind.
The idea is to start with a supersymmetric orbifold where all tachyonic states are projected out by the orbifold action itself, without any reference to
supersymmetry. (In the RNS language this means that all tachyons are projected out {\em before} the GSO projections.) This is achieved by
including non-trivial discrete torsion between some orbifold twists. We then break supersymmetry by including the Scherk-Schwarz
phase ($+1$/$-1$ when acting on space-time bosons/fermions) in some even order twists. Since this modifies the orbifold action only in space-time
{\em fermionic} sectors, the background remains tachyon-free.

{}While, for the sake of clarity, in this paper we focus on an explicit example, it would be interesting to systematically classify such orbifolds. Non-supersymmetric gauge theories on D3-branes in such orbifold backgrounds \cite{KS, LNV, BKV, BJ} is one potential application of our construction.
Thus, the absence of closed string tachyons at tree-level is encouraging as it appears to be necessary for non-supersymmetric
orbifold D3-brane gauge theories to have stable infrared fixed points \cite{BKV, TZ, CST, Fuchs, AS, DKR}.

{}The D3-brane gauge theory in our tachyon-free non-supersymmetric orbifold has a chiral ${\cal N} = 1$ supersymmetric spectrum but non-supersymmetric tree-level interactions. The 1-loop non-Abelian $\beta$-functions are vanishing in this gauge theory. We discuss 1-loop renormalization of quartic scalar couplings (double-trace operators). Since there are no tachyons, only massless twisted closed string states can contribute to this renormalization in the transverse channel. In non-supersymmetric twisted sectors we do have massless NS-NS states (while R-R states are all massive in these sectors). So, {\em a priori} double-trace operators can be renormalized at 1-loop. It would be interesting to see explicitly if there is a stable fixed point in the infrared.

{}Recently, tachyon-free non-supersymmetric Type II strings were discussed in \cite{Sch} in the context of rational conformal field theory
construction. Such backgrounds as well as free-fermionic \cite{KLT} and lattice \cite{LLS} models have compact (internal) directions.

\section{Non-Supersymmetric Orbifolds}\label{2}

{}The goal of this section is to describe a construction of non-compact tachyon-free non-supersymmetric orbifold
backgrounds. For definiteness, we will focus on Type IIB. Type IIA can be discussed similarly.

{}Consider Type IIB on ${\bf R}^{3,1} \times {\cal M}$,
where ${\cal M}$ is an orbifold. We can consider compact orbifolds ${\cal M} = T^6/\Gamma$,
where the orbifold group $\Gamma$ acts crystallographically on the 6-torus $T^6$.
For the reasons we discuss at the end of this section, we will consider non-compact orbifolds
${\cal M} = {\bf R}^6/\Gamma$ instead.

{}The orbifold group $\Gamma=\{g_a | a = 1, \dots, |\Gamma|\}$ ($g_1 = 1$) is
a finite discrete subgroup of $Spin(6)$. If $\Gamma\subset SU(3)(SU(2))$,
we have $1/4$ ($1/2$) of unbroken supersymmetries, and no supersymmetry, otherwise.

{}We will focus on cases where Type IIB on ${\cal M}$ is a modular invariant theory.
This is always the case if there are some unbroken supersymmetries. Otherwise,
this is also true if $\not\exists{\bf Z}_2\subset\Gamma$. If
$\exists{\bf Z}_2\subset\Gamma$, then modular invariance requires that the set of points in
${\bf R}^6$ fixed under the ${\bf Z}_2$ twist have real dimension 2.

{}The untwisted NS-NS and all R-R sectors contain massless and massive bosonic states. Twisted NS-NS sectors
contain massless and massive states in supersymmetric cases, while in non-supersymmetric cases may
also contain tachyons.

{}The strategy behind our construction is based on the following observations. In supersymmetric orbifolds tachyonic states in twisted sectors are automatically
projected out by the GSO projections. Imagine, however, that all tachyonic states are projected out by
the orbifold action itself {\em before} ({\em i.e.}, independently of)
the GSO projections. Then we can modify the orbifold action in the {\em fermionic sectors only} such
that supersymmetry is broken. The GSO projections are also modified, but the background remains tachyon-free as the absence of tachyons does not depend
on the GSO projections, and the orbifold action in the bosonic sectors is unaltered. We will now outline this construction in more detail.

{}Let $d_a$ be the real dimension of the $g_a$ fixed point locus in ${\cal M}$. Given orientability of $\Gamma$ we must have $d_a = 0,2,4$ for $a\not = 1$.
Twists with $d_a = 4$ break supersymmetry, and there are tachyons in the twisted sectors; will not include such twists.
Twists with $d_a = 2$ either preserve 1/2 of supersymmetries, or break them all (and in this case there are tachyons);
we will only include supersymmetric twists with $d_a = 2$. Twists with $d_a = 0$ can preserve 1/4 of
supersymmetries, or break them all. Such twisted sector states have no momentum modes in ${\cal M}$ but only oscillator excitations.
We will include certain non-supersymmetric twists with $d_a = 0$ (see below).

{}Our construction then goes follows. Consider an orbifold group $\Gamma \subset SU(3)$ with some $d_a = 0$.
It preserves 1/4 of supersymmetries. The ground state energy in the NS-NS $g_a$ twisted sectors with $d_a = 0$ is zero
if $n_a = 3,4$, where $n_a$ is the order of $g_a$, but is negative for some $g_a$ with $n_a > 4$. Our first
goal is to construct an orbifold where all tachyonic twisted states are projected out by the orbifold action itself
before the GSO projections.

{}This can be achieved by including discrete torsion corresponding to a non-trivial phase
in the action of $g_a$ on the $g_b$ twisted ground state \cite{Vafa}:
\begin{equation}\label{DT}
 g_a\left|0, g_b\right> = \exp(2\pi i\phi_{ab}) \left|0, g_b\right>~.
\end{equation}
Note that $\phi_{ad} + \phi_{bd} = \phi_{cd}$, where $g_c \equiv g_a g_b$.
Trivial discrete torsion $\phi_{ab} \equiv 0$ corresponds to geometric orbifolds. More generally, non-vanishing $\phi_{ab}$ are compatible with modular invariance.
For left-right symmetric orbifolds, which we focus on in this paper, we have (no summation over repeated indices):
\begin{eqnarray}
 &&\phi_{ab} = - \phi_{ba}~,\\
 &&n_a \phi_{ab} = 0~({\rm mod}~1)~.
\end{eqnarray}
Tachyonic states in the twisted sectors are not projected out by the orbifold action itself unless some $\phi_{ab} \neq 0$. So consider
an orbifold where for each twist $g_b$ with $d_b = 0$ and $n_b > 4$ we have a twist $g_a$ with $d_a = 2$ with $\phi_{ab} \neq 0$. Then
all tachyonic {\em ground} states in such $g_b$ twisted sectors are projected out. This does not guarantee absence of excited
tachyonic states; however, as we will see in the following, there exist cases where all excited states in such sectors are either massless or massive.
Note that since all twists with $d_a = 2$ by our construction
are supersymmetric, the corresponding twisted sectors are tachyon-free with or without discrete torsion.

{}Our second goal is to break supersymmetry by modifying the orbifold action in the fermionic sectors only.
This can be achieved as follows. Consider a twist $g_a \in \Gamma$ with even $n_a$. The action of $g_a$ is the same as before in the bosonic (NS-NS and R-R) sectors,
while it can be accompanied by a ${\bf Z}_2$ valued Scherk-Schwarz phase $\zeta_a$ in the fermionic (R-NS and NS-R) sectors\footnote{In the following we will always assume that $\zeta_a$ are the same in the left- and right-moving sectors, albeit in general this can be relaxed.}. In the GSO language,
our full orbifold group is $\Gamma \times {\bf Z}^L_2 \times {\bf Z}^R_2$, where the ${\bf Z}^L_2 \times {\bf Z}^R_2$
subgroup corresponds to the left- and right-moving GSO projections $G_L$ and $G_R$. The phase $\zeta_a$ is then nothing but
discrete torsion between $g_a$ and $G_L$ and $G_R$.
Such discrete torsion is consistent with modular invariance, which implies that the GSO projections are also accompanied by the same extra phase when acting on the $g_a$ twisted ground states:
\begin{eqnarray}\label{dt}
 G_{L,R} \left|0, g_a\right> = -\zeta_a (-1)^{F_{L,R}}\left|0, g_a \right>~,
\end{eqnarray}
where $F_L$ and $F_R$ are the left- and right-moving fermion number operators.

{}While the orbifold with trivial Scherk-Schwarz phases $\zeta_a$ preserves 1/4 of supersymmetries, in some cases including non-trivial $\zeta_a$ breaks all supersymmetries \cite{non-susy}. Examples include orbifolds with twists $g_a$ with $d_a = 0$ and odd $n_a$. The GSO projections no longer project out all twisted sector tachyons due to extra phases in (\ref{dt}). However, by our construction all tachyons are projected out by the orbifold action itself, so we obtain a modular invariant tachyon-free non-supersymmetric background. Systematic rules for orbifold model building,
which we use for bookkeeping in the following construction without giving details, can be found in \cite{KT}.

\subsection{A ${\bf Z}_3 \times {\bf Z}_2 \times {\bf Z}_2$ Example}\label{2.1}

{}Let the generators $g$, $R_1$ and $R_2$ of $\Gamma = {\bf Z}_3 \times {\bf Z}_2\times {\bf Z}_2$ act on the three complex coordinates
$z_i$ ($i = 1,2,3$) on ${\cal M} = {\bf C}^3 / \Gamma$ as follows ($R_3\equiv R_1 R_2$):
\begin{eqnarray}
 g : &&z_i \rightarrow \omega z_i~,\\
 R_i : &&z_j \rightarrow -(-1)^{\delta_{ij}} z_j~,
\end{eqnarray}
where $\omega \equiv \exp(2\pi i /3)$. The twist $g$ preserves 1/4 of supersymmetries. Individually, the twists $R_i$
preserve 1/2 of supersymmetries, while the ${\bf Z}_2\times {\bf Z}_2$ orbifold preserves 1/4 of supersymmetries. Without any discrete torsion,
the ${\bf Z}_3 \times {\bf Z}_2 \times {\bf Z}_2$ orbifold preserves 1/4 of supersymmetries.

{}Consider NS-NS twisted sectors. The (left- and right-moving) ground state energy in the $R_i$ as well as $g^{\pm 1}$ twisted sectors is 0.
The ground state energy in each of the $g^{\pm 1} R_i$ twisted sectors is $-1/6$ (in the units where the untwisted NS-NS ground state energy is $-1/2$).
In these sectors the lowest oscillator excitations contribute $1/6$ (see Subsection \ref{B2} of Appendix \ref{B} for details), so all excited states are either massless or massive. Our goal is to project out
tachyonic ground states before the GSO projections.

{}Consider non-trivial discrete torsion between $R_1$ and $R_2$:
\begin{eqnarray}\label{torsion}
 R_i\left|0, R_j\right> = -(-1)^{\delta_{ij}} \left|0, R_j\right>~,
\end{eqnarray}
which also implies that
\begin{eqnarray}
 R_i\left|0, g^{\pm 1} R_j\right> = -(-1)^{\delta_{ij}} \left|0, g^{\pm 1} R_j\right>~,
\end{eqnarray}
so all tachyonic ground states are projected out in the $g^{\pm 1} R_i$ twisted sectors, and they can only contain massless and massive states --
before the GSO projections.

{}The next step is to break supersymmetry by including non-trivial Scherk-Schwarz phases (\ref{dt}) in even order twists. Let
$\zeta_i$ be the phases corresponding to the twists $R_i$ ($\zeta_3 = \zeta_1\zeta_2$). Note that the same phases correspond to the twists $g^{\pm 1} R_i$.
With trivial phases $\zeta_i$ the orbifold preserves 1/4 of supersymmetries. However, if $\zeta_i$ are not all equal 1, then all supersymmetries are
broken\footnote{With asymmetric discrete torsion, which we do not consider here, the orbifold can also preserve 1/8 of supersymmetries.}.
To see this, let $N_i$ be the left-moving fermion number operators corresponding to complex fermions $\psi_i$, the
world-sheet fermionic superpartners of $z_i$. In the R-NS sector (the NS-R sector is analogous) the left-moving
ground state energy is zero, and only zero modes of $\psi_i$,
for which $N_i$ take values $0,1$, can be excited in massless states. For gravitinos, invariance under the twists $g$ and $R_i$ implies that (for more detail, see
Subsection \ref{B1} of Appendix \ref{B}):
\begin{eqnarray}
 &&{1\over 3}(N_1 + N_2 + N_3) = 0~({\rm mod}~1)~,\\
 &&{1\over 2}(N_2 + N_3) = \phi(\zeta_1)~({\rm mod}~1)~,\\
 &&{1\over 2}(N_1 + N_3) = \phi(\zeta_2)~({\rm mod}~1)~,
\end{eqnarray}
where $\phi(1) = 0$ and $\phi(-1) = 1/2$. We then have no gravitinos unless $\zeta_1 = \zeta_2 = 1$.

{}Non-trivial phases $\zeta_i$ break supersymmetry by modifying the orbifold action in the R-NS and NS-R sectors, but do not affect the orbifold action
in the bosonic sectors (albeit the GSO projections are modified). We therefore still have no twisted sector tachyons, which are projected out due to non-trivial discrete torsion (\ref{torsion}).

{}In Appendix A we discuss a compact version of this orbifold, where we do have twisted sector tachyons coming from fixed points away from the origin. These tachyons decouple in the decompactification limit, and this is one of the reasons why we focus on non-compact orbifolds. In Appendix B we discuss a bosonized RNS description, which provides a complementary way of seeing
how supersymmetry is broken and tachyons are projected out in this orbifold. By rotating a bosonic basis, we then discuss this orbifold in the light-cone Green-Schwarz description, both in bosonized and fermionic formulations.

\section{Adding D-branes}\label{3}

{}In Section \ref{2} we discussed closed strings on non-compact non-supersymmetric orbifolds without tachyons.
The goal of this section is to include open strings by adding
$N$ parallel D3-branes transverse to the orbifold ${\cal M}$.

{}We need to specify the action of the orbifold group on the Chan-Paton charges
carried by the D3-branes. It is described by $N \times N$ matrices $\gamma_a$ (corresponding to twists $g_a$),
which form a representation of $\Gamma$. Here $\gamma_1$ is the identity matrix $I_N$.

{}The open string one-loop vacuum amplitude is an annulus whose boundaries lie on D3-branes.
The elements $g_a\in \Gamma$ act on both ends of the open string, and this action corresponds to
$\gamma_a \otimes \gamma_a$ acting on the Chan-Paton indices. The annulus amplitude, therefore, has the
following form:
\begin{equation}\label{gsq}
 C = {1\over |\Gamma|}
 \sum_a \left({\rm Tr}(\gamma_a)\right)^2 {\widetilde C}_a~.
\end{equation}
The ``untwisted" contribution ${\widetilde C}_1$
is the same as in the ${\cal N} = 4$ theory for which
$\Gamma = \{1\}$. The information about the fact that the orbifold theory
has reduced (or broken) supersymmetry is encoded in the ``twisted" contributions ${\widetilde C}_a$, $a\not=1$.

\subsection{Tadpole and Anomaly Cancelation}\label{3.1}

{}From (\ref{gsq}) it follows that all tadpoles cancel if \cite{BKV}
\begin{equation}\label{trace}
 {\rm Tr}(\gamma_a) = 0~~~\forall~a \neq 1~.
\end{equation}
This also ensures absence of non-Abelian gauge anomalies in the D3-brane
gauge theory. Such anomalies would have to arise at the one-loop level. The relevant
diagram is the annulus with three external lines (with the Chan-Paton matrices $\lambda_1, \lambda_2, \lambda_3$)
corresponding to non-Abelian gauge fields attached to boundaries. First, consider a diagram with
two external lines attached to one boundary, and the third one to the other boundary.
The Chan-Paton structure of such a diagram is given by
\begin{equation}
 {\rm Tr}\left(\lambda_1\lambda_2\gamma_a\right){\rm Tr}\left(\lambda_3 \gamma_a\right)~,
\end{equation}
which vanishes because for non-Abelian gauge fields
${\rm Tr}(\lambda_r)=0$, so ${\rm Tr}(\lambda_r\gamma_a)=0$ ($\lambda_r$ are invariant under the orbifold
group action). Next, if all three external
lines are attached to one boundary, the corresponding Chan-Paton structure is given by
\begin{equation}\label{anomaly}
 {\rm Tr}\left(\lambda_1\lambda_2\lambda_3\gamma_a\right)
 {\rm Tr}\left(\gamma_a\right)~,
\end{equation}
which vanishes for $a\neq 1$ due to (\ref{trace}), hence no non-Abelian gauge anomalies.

\subsection{Chan-Paton Representations}\label{3.2}

{}In the absence of discrete torsion, the action of the orbifold group on the Chan-Paton
degrees of freedom is an $n$-fold copy of the regular representation of $\Gamma$ \cite{LNV, BKV}.
This $n$-fold copy satisfies (\ref{trace}), and we have $N = n |\Gamma|$.

{}If for ${\widetilde \Gamma} \subset\Gamma$ we have non-trivial discrete torsion (\ref{DT}), then
the corresponding Chan-Paton matrices form a projective representation of ${\widetilde \Gamma}$ \cite{Douglas}:
\begin{equation}
 \gamma_a\gamma_b = \exp(2\pi i\phi_{ab}) \gamma_c~,
\end{equation}
where $\gamma_c$ corresponds to $g_c = g_a g_b$. If ${\widetilde \Gamma} \neq \Gamma$,
we have a projective representation of ${\widetilde \Gamma}$ tensored with regular representations corresponding
to the other elements of $\Gamma$.

\subsection{D3-brane Gauge Theories}\label{3.3}

{}In the tachyon-free ${\bf Z}_3 \times {\bf Z}_2 \times {\bf Z}_2$ model of Section \ref{2} there is discrete torsion between $R_1$ and $R_2$
(\ref{torsion}). We must tensor copies of the regular representation of ${\bf Z}_3$ with copies of the (unique irreducible) projective
representation of ${\bf Z}_2\times {\bf Z}_2$ ($N = 12m$):
\begin{eqnarray}
 &&\gamma_g = {\rm diag}(1, \omega, \omega^{-1}) \otimes I_2 \otimes I_m~,\\
 &&\gamma_{R_i} = I_3 \otimes \sigma_i \otimes I_m~,
\end{eqnarray}
where $\sigma_i$ are Pauli matrices.
The D3-brane gauge theory {\em formally} has the massless spectrum of the ${\cal N} = 1$ supersymmetric $U(m)\times U(m)\times U(m)$ gauge theory with 3 copies of
chiral supermultiplets in $({\bf m}, {\bf \overline m}, {\bf 1})$, $({\bf 1}, {\bf m}, {\bf \overline m})$ and $({\bf \overline m}, {\bf 1}, {\bf m})$,
but with {\em non}-supersymmetric tree-level interactions\footnote{To see this spectrum note that in the ${\bf Z}_2\times {\bf Z}_2$ orbifold with discrete torsion \cite{Douglas} we have an ${\cal N} = 1$ supersymmetric $U(m)$ gauge theory with 3 copies of chiral supermultiplets in the adjoint, which is the same as the ${\cal N} = 4$ gauge theory spectrum, but the interactions are only ${\cal N} = 1$ supersymmetric. It then follows that the spectrum (but not the interactions) in the ${\bf Z}_3 \times {\bf Z}_2 \times {\bf Z}_2$ orbifold are the same as in the ${\bf Z}_3$ orbifold. The interactions are actually non-supersymmetric.}. The non-Abelian sector has vanishing 1-loop $\beta$-functions. Below we address the question of conformality beyond 1-loop.

{}One can also consider orientifold generalizations. Because of the ${\bf Z}_2$ twists, both O3 and O7 orientifold planes as well as D7-branes are present along with D3-branes. The D3-brane gauge theories in such orientifolds have non-vanishing 1-loop non-Abelian $\beta$-functions, which is due to the presence of the orientifold planes \cite{orient}.

\subsection{Is D3-brane Gauge Theory Conformal?}\label{4}

{}In \cite{BKV}, using the open string perturbative techniques, it was shown that planar diagrams for non-Abelian sectors of orbifold D3-brane gauge theories reduce to those in the parent ${\cal N} = 4$ gauge theory. The proof is based on vanishing of twisted planar diagrams with $b > 1$ boundaries and external lines corresponding to non-Abelian fields. This vanishing is due to the conditions (\ref{trace}). This implies that non-Abelian sectors of ${\cal N} = 2$ theories are finite in the large $N$ limit (and also for finite $N$). More generally, to see if a non-Abelian gauge theory is conformal, we need to check non-renormalization of: ({\em i}) gauge couplings; ({\em ii}) anomalous scaling dimensions; ({\em iii}) three-point Yukawa couplings; and ({\em iv}) quartic scalar couplings. Items ({\em iii}) and ({\em iv}) need to be checked only in non-supersymmetric cases as in ${\cal N} = 1$ theories we have a perturbative non-renormalization theorem for the superpotential. So non-Abelian sectors of ${\cal N} = 1$ theories are finite in the large $N$ limit \cite{BKV}\footnote{All $U(1)$ factors except for the center-of-mass $U(1)$ run and decouple in the infrared. To obtain conformal non-Abelian gauge theories, these $U(1)$ factors need to be removed \cite{BKV}. See \cite{DKR} for a detailed discussion.}.

{}However, in non-supersymmetric cases the situation is more subtle. In general, already at 1-loop the conformal property is spoiled. This is because of the quartic
scalar coupling renormalization \cite{BKV, TZ, CST, Fuchs, AS, DKR}. In the open string language the troublesome 1-loop diagram is an annulus with four external lines corresponding to scalars carrying non-Abelian charges attached to boundaries. Non-vanishing twisted diagrams then are those with two external lines attached to each boundary. These diagrams have the following Chan-Paton structure:
\begin{equation}\label{quartic}
 {\rm Tr}\left(\lambda_1\lambda_2\gamma_a\right){\rm Tr}\left(\lambda_3 \lambda_4\gamma_a\right)~,
\end{equation}
where $\lambda_r$ are Chan-Paton matrices corresponding to the scalars. Contracting external gauge indices, we obtain non-vanishing double-trace operators, which in general spoil conformality \cite{BKV, TZ, CST, Fuchs, AS, DKR}. In the transverse channel this is due to infrared divergences coming from tachyonic and massless closed string exchanges (while massive states in the transverse channel contribute finite terms to renormalization) \cite{BKV}. In \cite{DKR} it was shown that, at 1-loop, all tachyonic cases are non-conformal.

{}However, in our ${\bf Z}_3\times {\bf Z}_2\times {\bf Z}_2$ orbifold example we have no tachyons, so {\em a priori} there is a possibility that there is a stable infrared fixed point. Let us therefore discuss 1-loop renormalization of quartic scalar couplings in more detail. The twisted diagrams (\ref{quartic}) corresponding to twists $g_a$ that preserve some supersymmetries do no contribute to renormalization of quartic couplings. In our orbifold the twists $g^{\pm 1}$ as well as $R_i$ individually preserve 1/4 and 1/2 of supersymmetries, respectively (regardless of the values of $\zeta_i$), so we only need to consider the twists $g^{\pm 1} R_i$. Individually, these ${\bf Z}_6$ twists preserve 1/4 of supersymmetries if $\zeta_i = 1$, and break them all if $\zeta_i = -1$, so we need to focus on the twists $g^{\pm 1} R_i$ with $\zeta_i = -1$ (only two out of three $\zeta_i$ equal $-1$). Only closed string scalars propagate in the transverse channel, so we can focus on the NS-NS and R-R sector massless scalars (since we have no tachyons). We discuss the massless scalars in these twisted sectors in Subsection \ref{B2} of Appendix \ref{B}. There are no massless scalars in the R-R sectors, while NS-NS sectors contain massless scalars coming from left-right asymmetric world-sheet boson oscillator excitations (which is due to discrete torsion (\ref{torsion})). {\em A priori} these states can contribute to logarithmic renormalization of double-trace operators.

{}Here we note that in the supersymmetric case (all $\zeta_i = 1$) with discrete torsion (\ref{torsion}) all massless states in the NS-NS and R-R $g^{\pm 1}R_i$ twisted sectors are projected out, so these sectors do not contribute to logarithmic renormalization of double-trace operators. In other twisted sectors (as well as in the $g^{\pm 1}R_i$ twisted sectors in the supersymmetric case without discrete torsion) we have a cancelation between the NS-NS and R-R exchanges. In our non-supersymmetric ${\bf Z}_3\times {\bf Z}_2\times {\bf Z}_2$ orbifold example in non-supersymmetric $g^{\pm 1} R_i$ twisted sectors with $\zeta_i = -1$ we have massless NS-NS states but no massless R-R states. So, {\em a priori} double-trace operators can be renormalized at 1-loop. It would be interesting to see explicitly if there is still a stable fixed point in the infrared.

{}Finally, let us mention that massless closed string scalars in our tachyon-free non-supersymmetric backgrounds may develop tachyonic potentials beyond tree-level. Our point here is that in the case of freely-acting compact orbifolds tachyons appear in the small radius limit, while here we do not have any compact directions. Nonetheless, untwisted (such as the dilaton) and twisted massless scalars can still turn tachyonic at the loop level.

\subsection*{Acknowledgments}
{}I would like to thank Allan Adams for pointing out shortcomings in the original version of this paper. I am especially grateful to Cumrun Vafa for invaluable discussions on the current version.

\appendix

\section{Compact Cases}\label{A}

{}In this paper we have focused on non-compact orbifolds, where $g_a$ twists with $d_a = 0$ have only one fixed point at the origin of ${\cal M}$. In compact cases
the situation is more involved. Consider ${\cal M} = (T^2 \times T^2 \times T^2) / \Gamma$ with $\Gamma = {\bf Z}_3 \times {\bf Z}_2 \times {\bf Z}_2$
acting as above on the three complex coordinates $z_i$ parameterizing the three 2-tori. For definiteness, let us consider the case with $\zeta_1 = -1$, $\zeta_2 = -1$ (the cases with $\zeta_1 = -1$, $\zeta_2 = 1$ and $\zeta_1 = 1$, $\zeta_2 = -1$ are analogous). In the $g R_1$ twisted sector (the $g^{-1}R_1$ as well as $g^{\pm 1}R_2$ twisted sectors are analogous) there are 3 fixed points
in the first $T^2$, one of which is at the origin. Under the $R_2$ twist the fixed point at the origin is invariant, while the other two
are mapped into each other. One linear combination of the latter is invariant under $R_2$, while the other is a $-1$ eigenstate. This implies that even in the
presence of discrete torsion (\ref{torsion}), we have tachyonic states coming from the $g^{\pm 1} R_1$ twisted sectors from the fixed point combination which is odd under $R_2$. In the decompactification limit the twist fields corresponding to the two fixed points away from the origin decouple, and we are left with a tachyon-free non-compact orbifold we discussed in Subsection \ref{2.1}.

\section{Green-Schwarz Description}\label{B}

{}In this Appendix we discuss our ${\bf Z}_3 \times {\bf Z}_2 \times {\bf Z}_2$ orbifold in the light-cone Green-Schwarz description. We will do this in two steps. First, we bosonize the world-sheet fermions in the RNS formulation we discussed up until now. Then we rotate to a basis corresponding to bosonized spinors in the Green-Schwarz description, and then refermionize to arrive at the fermionic Green-Schwarz description.

\subsection{Bosonized RNS Description}\label{B1}
{}In the RNS formulation, in the light-cone gauge, we have eight real world-sheet fermions, which we complexify into four complex fermions
$\psi_I = (\psi_0, \psi_i)$, where $\psi_0$ is the world-sheet superpartner of the complex boson corresponding to the two real transverse directions in ${\bf R}^{3,1}$ (in the light-cone gauge), while $\psi_i$ ($i = 1,2,3$) are the world-sheet superpartners of the three complex bosons $z_i$ corresponding to
the three complex directions in ${\bf C}^3$ in ${\bf R}^{3,1} \times {\bf C}^3$. We can bosonize the complex fermions $\psi_I$ into
four real left- and right-moving chiral bosons $\varphi_I = (\varphi_0, \varphi_i)$. Oscillator excitations of $\varphi_I$ have integer modings. Momentum excitations are described by $SO(8)$ chiral lattices. Momentum states can be written as $\left|p_0, p_1, p_2, p_3\right>$, and have the energy contribution equal ${1\over 2} \sum_I (p_I)^2$. In this basis the vector ${\bf 8}_v$ of $SO(8)$ can be written as
\begin{eqnarray}
 \left|\pm 1, 0, 0, 0\right>, \left|0, \pm 1, 0, 0\right>, \left|0, 0, \pm 1, 0\right>, \left|0, 0, 0\pm 1\right>~,
\end{eqnarray}
while the spinors ${\bf 8}_s$ and ${\bf 8}_c$ can be written as
\begin{eqnarray}
 \left|\pm {1/ 2}, \pm {1 / 2}, \pm {1/ 2}, \pm {1/ 2}\right>~,
\end{eqnarray}
where the number of plus signs is even for ${\bf 8}_s$ and odd for ${\bf 8}_c$. Before orbifolding, the 1-loop Type IIB partition function can be written as:
\begin{equation}
 {\cal Z} = {1\over \eta^{12}(q)\eta^{12}({\overline q})} \sum_{p^L,p^R \in \chi} (-1)^{F_s}
 q^{{1\over 2} \left(p^L\right)^2} {\overline q}^{{1\over 2} \left(p^R\right)^2}~.
\end{equation}
Here $\chi$ is the chiral lattice corresponding to ${\bf 8}_v \oplus {\bf 8}_s$ plus descendants. Also,
\begin{equation}
 F_s \equiv 2(p^L_0 + p^R_0)~({\rm mod}~2)
\end{equation}
is the space-time fermion number operator: states with $F_s = 0$ are space-time bosons, while
states with $F_s = 1$ are space-time fermions.

{}The action of our ${\bf Z}_3 \times {\bf Z}_2 \times {\bf Z}_2$ orbifold on the bosonic coordinates $z_i$ is the same as before:
\begin{eqnarray}
 g : &&z_i \rightarrow \omega z_i~,\\
 R_i : &&z_j \rightarrow -(-1)^{\delta_{ij}} z_j~.
\end{eqnarray}
Its action on the bosons $\varphi_I$ is given in terms of {\em shifts} instead of twists:
\begin{eqnarray}
 g : &&\varphi_I \rightarrow \varphi_I + S(g)_I~,\\
 R_i : &&\varphi_I \rightarrow \varphi_I + S(R_i)_I~,
\end{eqnarray}
where
\begin{eqnarray}
 S(g) &=& (0, {1\over 3}, {1\over 3}, -{2\over 3})~,\\
 S(R_1) &=& (0, 0, {1\over 2}, -{1\over 2})~,\\
 S(R_2) &=& (0, -{1\over 2}, 0, {1\over 2})~,\\
 S(R_3) &=& (0, -{1\over 2}, {1\over 2}, 0)~.
\end{eqnarray}
We then have:
\begin{eqnarray}
 g\left|p^L\right> \otimes \left|p^R\right> &=& \exp\left[2\pi i S(g)\cdot \left(p^L - p^R\right)\right] \left|p^L\right> \otimes \left|p^R\right>~,\\
 R_i \left|p^L\right> \otimes \left|p^R\right> &=& \left(\zeta_i\right)^{F_s}
 \exp\left[2\pi i S(R_i)\cdot \left(p^L - p^R\right)\right] \left|p^L\right> \otimes \left|p^R\right>~,
\end{eqnarray}
where $\zeta_i$ are the ${\bf Z}_2$ valued Scherk-Schwarz phases ($\zeta_3 = \zeta_1\zeta_2$).
Finally, we also have non-trivial discrete torsion between the $R_1$ and $R_2$ twists:
\begin{eqnarray}\label{torsion.gs}
 R_i\left|0, R_j\right> = -(-1)^{\delta_{ij}} \left|0, R_j\right>~,
\end{eqnarray}
so $R_1$ acts with an extra (compared with the geometric orbifold action) minus sign on the $R_2$ twisted ground state.

{}The $R_i$ and $g^{\pm 1}$ twisted sectors have vanishing ground state energy, so they are tachyon-free. On the other hand, tachyons in the $g^{\pm 1} R_i$
twisted sectors are projected out. To see this, for definiteness, consider the $gR_1$ twisted sector (others are similar).
In this sector the ground state has the following $\varphi_I$ momenta (for both left- and right-movers):
\begin{equation}
 \left|0, {1\over 3}, -{1\over 6}, -{1\over 6}\right>~.
\end{equation}
This contributes $1/12$ into the ground state energy, while the twisted boson contribution is $1/4$, which gives the ground state energy of $-1/6$ (including the $-1/2$
vacuum contribution). First consider the $\varphi_I$ momentum descendants of the ground state. These descendants include the $SO(8)$ identity, vector as well as spinor descendants. The spinor descendants cannot be tachyonic. The identity descendants are all massive. The lowest vector descendant is of the form
\begin{equation}\label{gR_1}
 \left|0, -{2\over 3}, -{1\over 6}, -{1\over 6}\right>~,
\end{equation}
and contributes $1/4$ to the energy, so such states are at least massless. All oscillator excitations contribute at least additional $1/6$ (such $1/6$ excitations come from twisted $z_2$ and $z_3$ oscillators), so they are also all massless or massive. This means that only the ground state is tachyonic. However, (\ref{torsion.gs}) implies that
\begin{eqnarray}
 R_2\left|0, gR_1\right> = - \left|0, gR_1\right>~,
\end{eqnarray}
so the ground state is projected out by the $R_2$ twist\footnote{More precisely, this is the case in the non-compact case. As we discussed in Appendix A, in the compact case there is a projected-in tachyon corresponding to the $R_2$-odd linear combination of two $gR_1$ fixed points located away from the origin.}. Note that absence of tachyons does not rely on the GSO projections.

{}Finally, we need to show that supersymmetry is broken. Massless gravitinos of the unorbifolded theory come from the states ${\bf 8}_s \times {\bf 8}_v$ and ${\bf 8}_v \times {\bf 8}_s$. The orbifold breaks 10-dimensional Lorentz invariance $SO(9,1)$ to $SO(3,1) \times U(1)^3$. We can think about supersymmetry in terms of a 4-dimensional observer localized at the orbifold fixed point (or on D3-branes) located at the origin of ${\cal M}$. Then the corresponding gravitinos must have $p^R_0 = \pm 1$ and $p^L_0 = \pm 1$ for the ${\bf 8}_s \times {\bf 8}_v$ and ${\bf 8}_v \times {\bf 8}_s$ states, respectively. Such states are all projected out by the orbifold action as the $g$, $R_1$ and $R_2$ orbifold projections require that
\begin{eqnarray}
 &&{1\over 3}\left(p^L_1 + p^L_2 - 2p^L_3 - p^R_1 - p^R_2 + 2p^R_3\right) = 0~({\rm mod}~1)~,\\
 &&{1\over 2}\left(p^L_2 - p^L_3 - p^R_2 + p^R_3\right) = \phi(\zeta_1)~({\rm mod}~1)~,\\
 &&{1\over 2}\left(-p^L_1 + p^L_3 + p^R_1 - p^R_3\right) = \phi(\zeta_2)~({\rm mod}~1)~,
\end{eqnarray}
where $\phi(1) = 0$ and $\phi(-1) = 1/2$. We then have no gravitinos if either $\zeta_1 = -1$ or $\zeta_2 = -1$. So non-trivial discrete Scherk-Schwarz
phases $\zeta_i$ break supersymmetry by modifying the orbifold action on space-time fermions, but do not affect the orbifold action
in the bosonic sectors. Therefore, we still have no tachyons in the twisted sectors, which are projected out due to non-trivial
discrete torsion (\ref{torsion.gs}).

\subsection{Relation to RNS Free-Fermionic Description}\label{B2}

{}Let us give a translation between the bosonized and the RNS free-fermionic descriptions. Let us focus on left-movers (right-movers are similar).
Consider momentum states $\left|p_0,p_1,p_2,p_3\right>$. Here $p_0$ can take only integer values (NS sectors) or half-odd-integer values (R sectors), and
the $\psi_0$ fermion modings are then half-odd-integer and integer, respectively. The values of $p_i$ momenta depend on the twisted sectors. The $\psi_i$
fermion modings are given by $\pm(1/2 - p_i)~({\rm mod}~1)$. For example, in the NS-NS $gR_1$ twisted sector the ground state in the bosonized language is
given by (\ref{gR_1}). In the free-fermionic language the creation operators for the fermion oscillators are given by ($n\in {\bf N}$):
\begin{eqnarray}
 &&\psi_0 : b^\dagger_{n - 1/2}~,~~~d^\dagger_{n - 1/2}~,\\
 &&\psi_1 : b^\dagger_{n - 5/6}~,~~~d^\dagger_{n - 1/6}~,\\
 &&\psi_2 : b^\dagger_{n - 1/3}~,~~~d^\dagger_{n - 2/3}~,\\
 &&\psi_3 : b^\dagger_{n - 1/3}~,~~~d^\dagger_{n - 2/3}~.
\end{eqnarray}
The corresponding twisted boson oscillator modes are given by (the world-sheet superpartner of $\psi_0$ is integrally moded):
\begin{eqnarray}
 &&z_1 : b^\dagger_{n - 2/3}~,~~~d^\dagger_{n - 1/3}~,\\
 &&z_2 : b^\dagger_{n - 1/6}~,~~~d^\dagger_{n - 5/6}~,\\
 &&z_3 : b^\dagger_{n - 1/6}~,~~~d^\dagger_{n - 5/6}~.
\end{eqnarray}
In this language we can see that in the $gR_1$ twisted sector only the NS-NS ground state is tachyonic with energy $-1/6$, while all excited states are at least massless as the lowest oscillators contribute $1/6$. The tachyonic NS-NS ground states as well as the massless R-R ground states in the $gR_1$ twisted sector are
projected out due to non-trivial discrete torsion (\ref{torsion.gs}). Before the GSO and $R_2$ projections, the massless states in the NS-NS $gR_1$ twisted sectors come from the left- and right-moving oscillator excitations (with 1/6 moding) corresponding to $\psi_1$, $z_2$ and $z_3$. The GSO projections project out the states with $\psi_1$ excitations (here we are assuming $\zeta_1 = -1$; if $\zeta_1 = 1$, then the $gR_1$ twist is supersymmetric, and $\psi_1$ excitations are projected in instead). The $R_2$ projection keeps only two massless states: with the $z_2$ left-moving and the $z_3$ right-moving excitations, and with the $z_3$ left-moving and the $z_2$ right-moving excitations. Note that these states have left-right asymmetric oscillator excitations.

\subsection{Bosonized Green-Schwarz Description}\label{B3}

{}In the RNS description the world-sheet fermions $\psi_I$
transform in the vector representation ${\bf 8}_v$ of $SO(8)$. In the Green-Schwarz formulation the fermions transform
in a spinor representation. Since above we chose the ground state to be $\left|{\bf 8}_v\right>\oplus\left|{\bf 8}_s\right>$, the spinor representation for the
fermions in the Green-Schwarz description is ${\bf 8}_c$. We can obtain these fermions by rotating the bosons $\varphi_I$ (corresponding to $\psi_I$) from the
vector basis ${\bf 8}_v$ to the spinor basis ${\bf 8}_c$, and then refermionizing. The bosons in the ${\bf 8}_c$ basis read:
\begin{eqnarray}
 &&\rho_0 = {1\over 2}\left(-\varphi_0 + \varphi_1 + \varphi_2 + \varphi_3 \right)~,\\
 &&\rho_1 = {1\over 2}\left(-\varphi_0 + \varphi_1 - \varphi_2 - \varphi_3 \right)~,\\
 &&\rho_3 = {1\over 2}\left(-\varphi_0 - \varphi_1 + \varphi_2 - \varphi_3 \right)~,\\
 &&\rho_4 = {1\over 2}\left(-\varphi_0 - \varphi_1 - \varphi_2 + \varphi_3 \right)~.
\end{eqnarray}
Note that the orbifold action on $\rho_A$, $A = 0,1,2,3$ is identical to its action on $\varphi_I$, $I = 0,1,2,3$:
\begin{eqnarray}
 g : &&\rho_A \rightarrow \rho_A + V(g)_A~,\\
 R_i : &&\rho_A \rightarrow \rho_A + V(R_i)_A~,
\end{eqnarray}
where
\begin{eqnarray}
 V(g) &=& (0, {1\over 3}, {1\over 3}, -{2\over 3})~,\\
 V(R_1) &=& (0, 0, {1\over 2}, -{1\over 2})~,\\
 V(R_2) &=& (0, -{1\over 2}, 0, {1\over 2})~,\\
 V(R_3) &=& (0, -{1\over 2}, {1\over 2}, 0)~.
\end{eqnarray}
This is because of the triality symmetry of $SO(8)$ (under which ${\bf 8}_v \rightarrow {\bf 8}_s \rightarrow {\bf 8}_c \rightarrow {\bf 8}_v$). So all of the
discussion of Subsection \ref{B1} can be ported over into this basis completely unchanged. In particular, we have the following action
($p^L_A$ and $p^R_A$ are the left- and right-moving momenta corresponding to the bosons $\rho_A$):
\begin{eqnarray}
 g\left|p^L\right> \otimes \left|p^R\right> &=& \exp\left[2\pi i V(g)\cdot \left(p^L - p^R\right)\right] \left|p^L\right> \otimes \left|p^R\right>~,\\
 R_i \left|p^L\right> \otimes \left|p^R\right> &=& \left(\zeta_i\right)^{F_s}
 \exp\left[2\pi i V(R_i)\cdot \left(p^L - p^R\right)\right] \left|p^L\right> \otimes \left|p^R\right>~,
\end{eqnarray}
where $\zeta_i$ are the ${\bf Z}_2$ valued Scherk-Schwarz phases ($\zeta_3 = \zeta_1\zeta_2$).
Finally, we also have non-trivial discrete torsion (\ref{torsion.gs}).

\subsection{Fermionic Green-Schwarz Description}\label{B4}

{}We can now refermionize the bosons $\rho_A$ to obtain four complex fermions $\chi_A$, which are a complexification of eight real fermions $S_{\dot a}$ transforming in ${\bf 8}_c$ of $SO(8)$. The orbifold action on $\chi_A$ is given by ($g$ and $R_i$ act trivially on $\chi_0$):
\begin{eqnarray}
 g &:& \chi_i \rightarrow \omega \chi_i~,\\
 R_i &:& \chi_j \rightarrow -(-1)^{\delta_{ij}} \chi_j~,
\end{eqnarray}
which determines the modings of $\chi_A$ (as in Subsection \ref{B2}).
The action of twists $R_i$ is accompanied by Scherk-Schwarz phases $\zeta_i$ when acting on the spinor ground states. This can be described as follows.
For our purposes it will suffice to consider only the $R_1$ twist. First consider the twist ${\widetilde R}_1$ corresponding to the geometric ${\bf Z}_2$
orbifold ({\em i.e.}, with a trivial Scherk-Schwarz phase.). It breaks $SO(8)$ to $SO(4)^2\approx SU(2)^4$. Under this twist all states are even except for ${\bf 2}$ of {\em one} of the four $SU(2)$ subgroups, which is odd. (This action can be represented as a $1/2$ shift $\sqrt{2}/2$ in the corresponding $SU(2)$ root lattice.)
Without loss of generality we can choose it to be the last $SU(2)$. We have the following decomposition of the $SO(8)$
representations under the branching $SO(8)\supset SU(2)^4$:
\begin{eqnarray}
 {\bf 1} &=& ({\bf 1}, {\bf 1}, {\bf 1}, {\bf 1}) + ({\bf 2}, {\bf 2}, {\bf 2}, {\bf 2}_-)~,\\
 {\bf 8}_v &=& ({\bf 2}, {\bf 2}, {\bf 1}, {\bf 1}) + ({\bf 1}, {\bf 1}, {\bf 2}, {\bf 2}_-)~,\\
 {\bf 8}_s &=& ({\bf 1}, {\bf 2}, {\bf 2}, {\bf 1}) + ({\bf 2}, {\bf 1}, {\bf 1}, {\bf 2}_-)~,\\
 {\bf 8}_c &=& ({\bf 2}, {\bf 1}, {\bf 2}, {\bf 1}) + ({\bf 1}, {\bf 2}, {\bf 1}, {\bf 2}_-)~,
\end{eqnarray}
where the subscript in ${\bf 2}_-$ indicates that it is odd under ${\widetilde R}_1$. This implies that the action of ${\widetilde R}_1$ on the ground states is given by:
\begin{eqnarray}
 &&{\bf 8}_v : {\widetilde R}_1 \left|{\bf 2}, {\bf 2}, {\bf 1}, {\bf 1}\right> = \left|{\bf 2}, {\bf 2}, {\bf 1}, {\bf 1}\right>,~~~
 {\widetilde R}_1 \left|{\bf 1}, {\bf 1}, {\bf 2}, {\bf 2}_-\right> = -\left|{\bf 1}, {\bf 1}, {\bf 2}, {\bf 2}_-\right>,\\
 &&{\bf 8}_s : {\widetilde R}_1 \left|{\bf 1}, {\bf 2}, {\bf 2}, {\bf 1}\right> = \left|{\bf 1}, {\bf 2}, {\bf 2}, {\bf 1}\right>,~~~
 {\widetilde R}_1 \left|{\bf 2}, {\bf 1}, {\bf 1}, {\bf 2}_-\right> = -\left|{\bf 2}, {\bf 1}, {\bf 1}, {\bf 2}_-\right>.
\end{eqnarray}
Now we can describe the action of $R_1$ as follows:
\begin{eqnarray}
 &&{\bf 8}_v : R_1 \left|{\bf 2}, {\bf 2}, {\bf 1}, {\bf 1}\right> = \left|{\bf 2}, {\bf 2}, {\bf 1}, {\bf 1}\right>,~~~
 R_1 \left|{\bf 1}, {\bf 1}, {\bf 2}, {\bf 2}_-\right> = -\left|{\bf 1}, {\bf 1}, {\bf 2}, {\bf 2}_-\right>,\\
 &&{\bf 8}_s : R_1 \left|{\bf 1}, {\bf 2}, {\bf 2}, {\bf 1}\right> = \zeta_1\left|{\bf 1}, {\bf 2}, {\bf 2}, {\bf 1}\right>,~~~
 R_1 \left|{\bf 2}, {\bf 1}, {\bf 1}, {\bf 2}_-\right> = -\zeta_1\left|{\bf 2}, {\bf 1}, {\bf 1}, {\bf 2}_-\right>.
\end{eqnarray}
Note that the twist $R_1$ preserves 1/2 of supersymmetries for either value $\zeta_1 = \pm 1$.

{}Next, consider the action of $g$. By itself it breaks $SO(8)$ to $SO(2)\times U(3)$. Together with $R_1$ the breaking is $SO(2) \times U(1) \times U(2)$.
The surviving $SU(2)$ is precisely the last $SU(2)$ in the above $SU(2)^4$ basis.
(The action of $g$ can be represented as the shift $(1/3\sqrt{2}, 1/3\sqrt{2}, -2/3\sqrt{2}, 0)$ in the
$SU(2)^4$ lattice.) This implies that $g$ preserves 1/4 of supersymmetries, and the corresponding gravitinos come from (the appropriate subset of)
$\left|{\bf 1}, {\bf 2}, {\bf 2}, {\bf 1}\right>$. These states survive the $R_1$ projection if $\zeta_1 = 1$, but for a non-trivial Scherk-Schwarz phase
$\zeta_1 = -1$ no gravitinos survive.

{}Note that the $R_2$ twist is not needed to break supersymmetry, but to project out tachyons in the $g^{\pm 1} R_i$ twisted sectors
via non-trivial discrete torsion (\ref{torsion.gs}).


\end{document}